\documentclass[english]{sig-alternate}
\usepackage[T1]{fontenc}
\usepackage[latin9]{inputenc}
\usepackage{graphicx}

\makeatletter
\@ifundefined{date}{}{\date{}}
\pdfoutput=1 
\makeatletter
\def\@copyrightspace{\relax}
\makeatother

\numberofauthors{3}
\author{
\alignauthor Teresa Leyk\\ 
\affaddr{Department of Computer Science and Engineering}\\
\affaddr{Texas A\&M University}\\
\affaddr{College Station, TX 77843}\\
\email{teresa19@tamu.edu}
\alignauthor Robert McInvale\\ 
\affaddr{Department of Computer Science and Engineering}\\
\affaddr{Texas A\&M University}\\
\affaddr{College Station, TX 77843}\\
\email{rrmcinv@gmail.com}
\alignauthor Ling Chen\\
\affaddr{Department of Computer Science and Engineering}\\
\affaddr{Texas A\&M University}\\
\affaddr{College Station, TX 77843}\\
\email{lciis@tamu.edu}
}

\makeatother

\usepackage{babel}
\begin{document}
\selectlanguage{english}

\title{Structured Peer Learning Program \textendash{} An Innovative Approach
to Computer Science Education}

\subtitle{Case Study}
\maketitle
\begin{abstract}
Structured Peer Learning (SPL) is a form of peer-based supplemental
instruction that focuses on mentoring, guidance, and development of
technical, communication, and social skills in both the students receiving
assistance and the students in teaching roles. This paper explores
the methodology, efficacy, and reasoning behind the practical realization
of a SPL program designed to increase student knowledge and success
in undergraduate Computer Science courses. Students expressed an increased
level of comfort when asking for help from student teachers versus
traditional educational resources, historically showed an increased
average grade in lower-level courses, and felt that the program positively
impacted their desire to continue in or switch to a Computer major.
Additionally, results indicated that advances in programming, analytical
thinking, and abstract analysis skills were evident in not only the
students but also the student teachers, suggesting a strong bidirectional
flow of knowledge.
\end{abstract}


\section{Introduction }

Compared to Biology, Chemistry, or Physics, the discipline of Computer
Science is relatively new, having emerged into the public consciousness
as late as the mid-1900s, see \cite{Den00}. However, the Computer
Science job market is consistently strong, and Computer Systems Analyst
and Software Developer jobs have been ranked at the top of the list
of \textquotedblleft The 100 best jobs\textquotedblright{} by the
US News \& World Report, see \cite{LRK14}. Unfortunately, due to
the specific educational demands of the subject, this job market success
does not automatically translate into success for students studying
Computer Science and Engineering. Computer Science requires a unique
mode of thought, emphasizing critical thinking, extensive analysis,
and a degree of creativity, and also requires from students the ability
to express themselves using the very strict syntax of code. Much like
writing in English or a foreign language, students must learn the
intricacies of using a programming language, and as many as 40\% of
students entering the Computer Science major have never seen a line
of code before, see \cite{Str09}.

These challenges can be difficult for Computer Science educators to
overcome. Not only must students be taught a plethora of abstract
concepts, but they require the experience and knowledge necessary
to concretely implement them. To this end, instructors have employed
a number of traditional methods in an attempt to establish and then
reinforce concepts in the minds of their students. These methods often
focus on one-directional learning (lecture) and the issuance of reinforcing
assignments (homework), and while these approaches are linchpins in
any successful learning experience, some of their more outstanding
drawbacks have been shown to have an especially deleterious impact
on students of Computer Science. Lectures and homework have a tendency
to place students in passive roles, to emphasize one-way communication,
and to require a large amount of unguided time outside of class. This
can lead to students improperly reinforcing what they have learned
which can lead to the repetition of mistakes and misconceptions. Furthermore,
these methods can be distinctly unsuited to teaching complex abstract
material, see \cite{Bon96}.\enlargethispage{\baselineskip}

These weaknesses in traditional methods have led to the creation of
several supplementary means of instruction intended to engage students
in a more active way, see \cite{Pri04}, including the concept of
SPL. SPL is similar to Peer Tutoring, which is a form of supplementary
instruction in which people from similar social groupings, who are
not professional teachers, help each other to learn and learn themselves
by teaching, see \cite{Fal01,Gol76,Top96}. SPL differs from Peer
Tutoring in that the system is formalized (with students being hired
by and working directly for the department), and in SPL emphasis is
on bidirectional learning and development of not only the students
who are receiving aid, but of the student teachers as well. This system
is capable of addressing many of the weak points in traditional learning,
acting as a supplement to more passive methods. By providing a resource
to students comprised of their peers, students are more able to engage
with the material they are learning, and are more comfortable while
doing so, see \cite{NWi66}. Students are encouraged to participate
actively, communication becomes bidirectional, and mistakes can be
corrected before they become reinforced. Finally, SPL offers students
a relaxed learning environment in which they can ask questions without
feeling the pressure of authority, increasing the number of students
seeking assistance as well as improving the learning experience of
those who do.

\section{Structured Peer Learning}

The Computer Science and Engineering (CSE) department at Texas A\&M
University established its structured peer learning (SPL) program
with the goal of increasing retention in CSE, with the additional,
related objective of increasing student success in Computer Science
courses. The program is staffed by high-performing students who are
hired part-time and assigned specific Computer Science classes that
correlate to their proficiencies (generally based on prior outstanding
performance in the courses they are assigned to teach). As these students
graduate and leave the program, they are replaced with new high-performing
students, generally selected from the lower-level undergraduate body.
In this way, new student teachers are constantly cycled through and
afforded the chance to learn from the opportunities provided by the
program, and the relevancy of student teacher classroom experience
is kept current by ensuring that student teachers assigned to a specific
class have recently taken that class.

Student teachers are compensated by the department, are required to
complete an initial training regimen, and are required to attend training
sessions every semester for purposes of professionalization. All student
teachers must abide by a code of conduct and a set of rules established
by the university and the CSE department, and are expected to represent
the school in a dignified, professional manner.

\subsection{Benefits to Students}

There are several ways in which student teachers provide assistance
to students:
\begin{itemize}
\item Assisting, tutoring, and instructing directly in Computer Science
course labs. This allows students to address issues which are fresh
in the minds of students, and to provide direct assistance as students
are working on completing assignments.
\item Providing dedicated hours during which students can approach student
teachers for one-on-one tutoring. Student teachers also enroll in
any electronic resources (message boards or forums) tied to the class,
and regularly respond to questions posed by students on these media.
\item Leading supplemental classroom-based instruction for groups of students.
These sessions are generally designed to review concepts presented
in lecture, and are oriented towards reiterating points so that students
strengthen their grasp on course material.
\end{itemize}
It is an important tenant of a SPL program that certain formal aspects
of traditional instruction, such as curriculum development and grading,
are not performed by student teachers, so as not to blur the boundary
between student and instructor, see \cite{RHT01}. The program grants
student teachers a degree of space from the authority figures leading
a course:
\begin{itemize}
\item Grading is never performed by student teachers. This is important
to ensure that a peer dynamic is preserved, as formal evaluation can
easily disrupt any feeling of equal footing.
\item Student teachers generally collaborate with the teaching assistants
(TAs) heading labs, but they usually do not take immediate direction
from instructors. 
\item Formal classroom direction is not disseminated through student teachers.
Students do not receive assignments from student teachers, and student
teachers are not able to instruct students to perform course objectives.
This is very important in maintaining a peer relationship between
student teachers and students.
\end{itemize}
We believe that this space between student teacher and formal instructor
is a necessary condition for the unique dynamic that develops between
students and their student teachers. Instead of being formally empowered
with authority in the course or passing down mandates from ``on high'',
student teachers are able to act as relatable resources for their
students, and are able to provide insight not only into technical
matters, but also into the social, cognitive, and circumstantial factors
surrounding an undergraduate career that includes a Computer Science
curriculum.

\subsection{Benefits to Student Teachers}

A critical component of SPL, and one that differentiates it from Peer
Tutoring, is the vast number of benefits enjoyed by student teachers
both while they remain part of the program, and after they have graduated
and moved into further education or the workforce. Many of these benefits
are academic:
\begin{itemize}
\item Student teachers are required to repeatedly teach complex Computer
Science topics to students, which serves to cement an understanding
of those concepts in the student teachers' minds. While student teachers
are generally already high-performing students, this repetition helps
to promote further mastery of many topics that are central to their
success in the field.
\item Computer Science is a field in which many problems can be solved in
multiple ways, and as such student teachers are often exposed to alternate
solutions when assisting students with assignments. This encourages
a flexible mode of thought and a broad range of thinking, which helps
to expand the skill set of student teachers.
\item Because different professors may teach the same course differently
and student teachers generally assist any and all students in a particular
course, regardless of professor, student teachers are exposed to multiple
facets of the subjects they teach, which broadens their perspectives
and allows for a more complete education on the topic.
\end{itemize}
In addition to these academic benefits, student teachers are exposed
to many advantageous scenarios which may increase their success in
the workforce and/or graduate school:
\begin{itemize}
\item In the implementation of SPL in the CSE department at Texas A\&M University,
student teachers assist in running the SPL program, and undertake
many projects which provide support, directly or indirectly, to the
department. These projects range from managing scheduling and assisting
with hiring, to website maintenance and construction, to the development
of applications and interfaces that facilitate the success of other
student teachers. These projects and responsibilities cultivate valuable
skills and experience in the student teaching staff, and can also
help to strengthen student teacher resumes.
\item By interacting in a work environment with one another, student teachers
build a network of high-performing peers that may carry forward into
their future careers.
\item Student teachers are well-positioned to mentor not only their students,
but each other as well. Valuable experience gained during internships,
co-ops, or employment, as well as impressions from courses taken,
can be shared by more senior student teachers with their juniors.
\end{itemize}
These benefits can greatly impact student teachers, and serve to provide
distinct benefits to the active student body that may not be seen
in other methods of education, such as traditional Peer Tutoring,
see \cite{CLi11}.

\section{Historical Data}

The Computer Science and Engineering (CSE) department at Texas A\&M
University performed research in 2007-{}-2008 on the effectiveness
of the SPL program as it pertains to student academic success, the
results of which were submitted to the Texas Higher Education Coordinating
Board but not formally published. Students were surveyed to determine
whether they asked questions of student teachers during the semester,
and their GPAs (grade point averages) were assessed at the end of
their courses. The results indicated significant improvement in the
academic performance of underclassmen who took advantage of the SPL
program, with less conclusive results for students in upper level
(junior and senior level) courses.

\subsection{Methodology}

Approximately 1100 students taking Computer Science courses were surveyed
as part of the study, with about 500 responding (approx. 200 of whom
were in freshman level courses, approx. 100 in sophomore level, and
approx. 200 junior and senior level). Information was sought from
students whether or not they asked Student Teachers any questions
throughout the semester, Those students course grades at the end of
the semester were recorded. Grades were then converted to a 4-point
scale and averaged by course level.

\subsection{Results}

Students in Freshman and Sophomore level classes showed a clear increase
in grade earned (Fig. 1), with students in freshman classes who asked
questions of Student Teachers during the semester earning an average
grade of 2.77, while those who did not ask questions of Student Teachers
earning an average grade of 2.09, a difference of 0.68 points. Students
in sophomore level courses were also greatly benefited, with students
who asked questions earning an average grade of 2.95 and students
who did not earning a grade of 2.36, a difference of 0.59 points.
The benefits to students in junior and senior level courses were less
clear (2.96 questions, 2.95 no questions). Furthermore, 65.48\% of
the students who asked questions ended up ``succeeding'' (defined
as earning either an A or a B grade) in their courses, whereas only
58.34\% of students who did not ask questions ended up succeeding
(Fig. 2). 

Finally, students were surveyed at the end of semester to evaluate
their level of comfort in asking questions of instructors, teaching
assistants (TAs) and student teachers (Fig. 3). In freshman classes,
a greater proportion of students indicated that they were comfortable
asking questions of student teachers (87\%) than indicated that they
were comfortable asking questions of TAs (83\%) or instructors (74\%).
Similarly, more sophomore students felt comfortable asking questions
of student teachers (91\%) than did those who felt comfortable asking
questions of TAs (89\%) or instructors (62\%). In upper level courses,
students were most comfortable asking questions of instructors (92\%),
followed by student teachers (88\%) and TAs (68\%).
\begin{center}
\begin{figure}[h]
\begin{centering}
\includegraphics[scale=0.6]{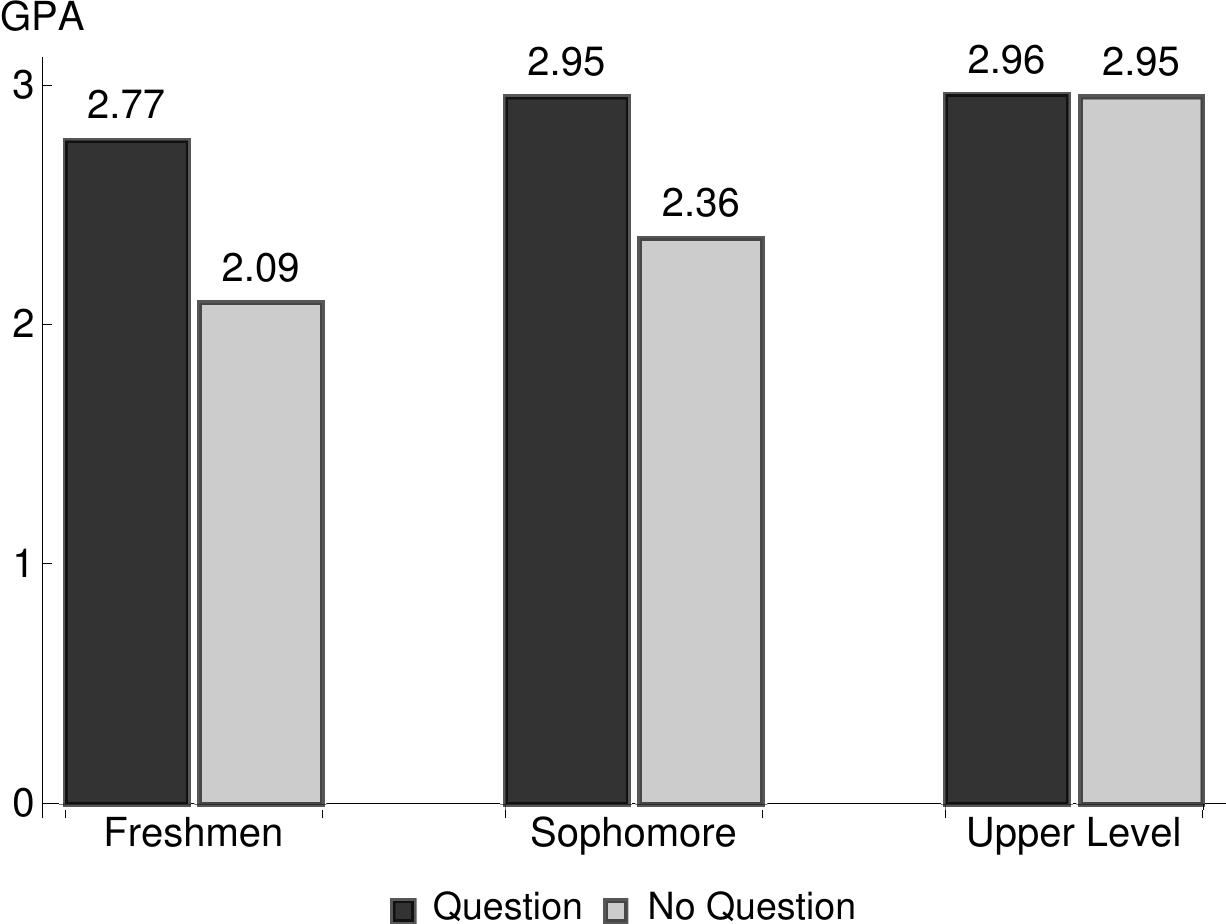}
\par\end{centering}
\centering{}\caption{Student GPA: Questions vs No Questions (Historical Data)}
\end{figure}
\par\end{center}

\begin{center}
\begin{figure}[h]
\begin{centering}
\includegraphics[scale=0.6]{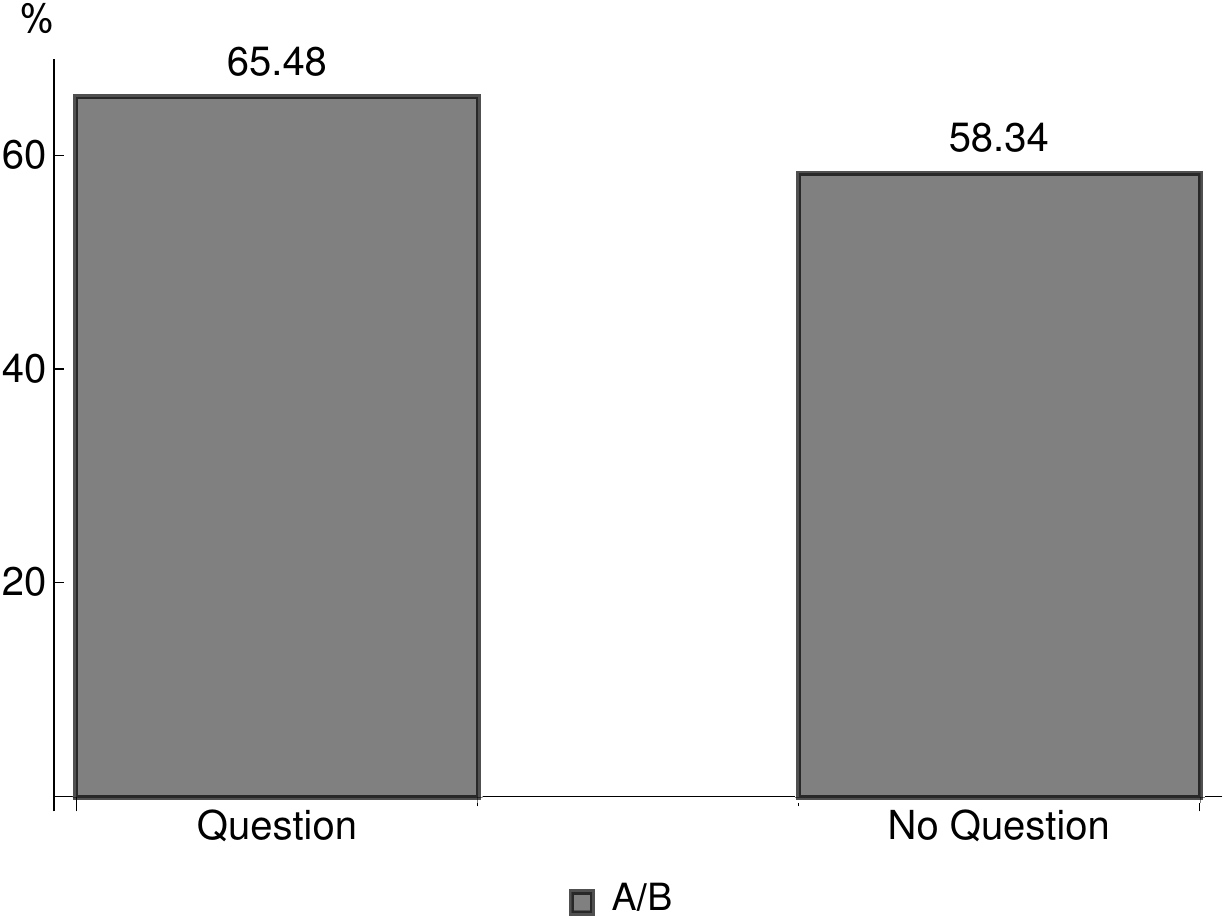}
\par\end{centering}
\centering{}\caption{Student Success: Questions vs No Questions (Historical Data)}
\end{figure}
\par\end{center}

\begin{center}
\begin{figure}[h]
\begin{centering}
\includegraphics[scale=0.6]{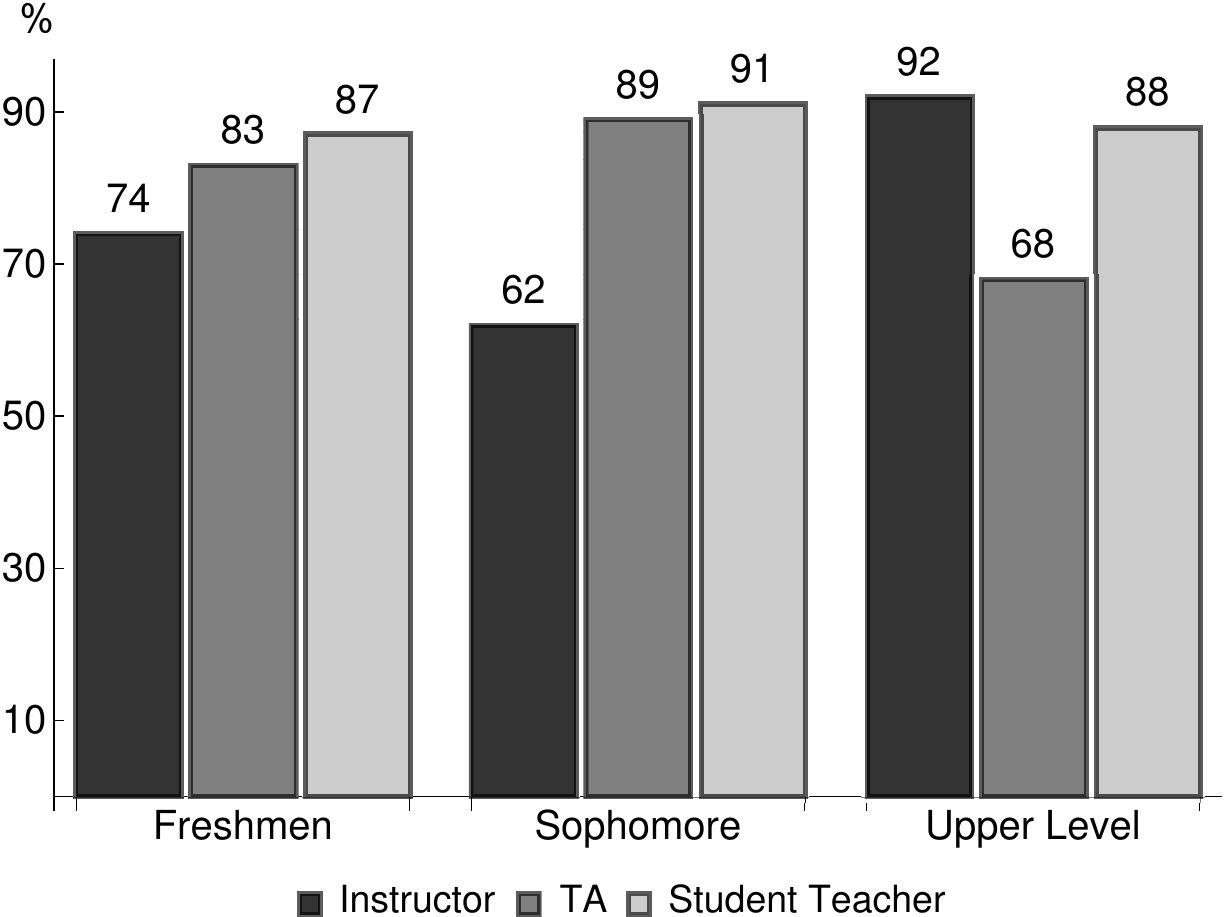}
\par\end{centering}
\centering{}\caption{Level of Comfort (Historical Data)}
\end{figure}
\par\end{center}

\section{Research Objectives}

The SPL program in the CSE department at Texas A\&M University was
established as a means to increase retention in-major, but it has
also proven an excellent testbed for evaluating the potential benefits
of SPL itself. The program was not designed as a rigorous experiment,
and no controls were implemented as part of this program, as the initial
and ongoing goal has been to help as many students as possible and
not necessarily to measure success with exacting rigor. Therefore,
many of  research objectives rely on the subjective experiences of
the students and student teachers involved.

With this in mind, we set out to measure several key areas of student
growth, with a focus on evaluating development of both the student
and the student teachers. We hoped to answer the following questions: 
\begin{itemize}
\item What educational benefits are received by students who seek help?
\item To what extent were student teachers able to deepen their computer
science knowledge?
\item What, if any, is the environmental impact of SPL? Are students more
comfortable in seeking help from student teachers as opposed to other,
more conventional sources? 
\item What are the fringe benefits of SPL? Do student teachers see significant
improvement in non-technical skills that are conducive to success
within the field of Computer Science?
\end{itemize}

\section{Methodology}

We collected data primarily through means of online and written surveys.
At the end of each Computer Science course in which student teachers
played an instructive role, students of the course were asked to complete
a short questionnaire evaluating the student teachers they interacted
with. Additionally, former and graduating student teachers were contacted
and asked to complete a survey evaluating the efficacy and impact
of the SPL program on their college and/or professional careers.

Student survey data was collected beginning in spring of 2009 and
continued through spring 2016. During this time we received and analyzed
a total of 4709 responses. Questions encompassed student level of
comfort in asking questions of student teachers vs traditional instructors,
impact of SPL on student understanding and course grades, impact of
student teachers on CSE retention, and effectiveness of the student
teachers in general. 

Former student teacher survey data was collected in spring of 2016.
We received 33 responses from students who worked as student teachers
at some point during the period from 2007 to 2016. Questions encompassed
overall satisfaction with the SPL program, personal benefits attained
from participation in the program, and the professional and academic
impact of serving as a student teacher. Many of the questions asked
were descriptive multiple choice, but respondents were also encouraged
to provide comments and suggestions for the program in free response
form.

\section{Results}

Students were asked to rate their level of comfort with asking questions
of student teachers, teaching assistants (TAs), and course instructors
by evaluating the statement \textquotedblleft \emph{I am comfortable
asking \_\_\_\_ when I do not understand a topic being discussed in
class.}\textquotedblright{} Responses were rated on a 4-point scale
(4\textendash strongly agree, 3\textendash agree, 2\textendash disagree,
1\textendash strongly disagree). Students rated their comfortability
level with student teachers (Fig. 4) at an average of 3.63 (standard
deviation 0.54), displaying a consistently higher level of comfortability
than with TAs (3.36, standard deviation 0.69) or instructors (3.32,
standard deviation 0.69). Additionally, students were much more likely
to rate their level of comfort with student teachers at the maximum
possible score, with 70.8\% of responses falling under \textquotedblleft strongly
agree\textquotedblright , compared with 52.4\% and 48.4\% for TAs
and instructors, respectively. 

Student responses were overwhelmingly positive regarding the impact
of the program on academic success: 92.0\% of respondents agreed that
the program helped to improve their understanding of the course material,
with 89.5\% indicating that their grades improved as a result of the
program (Fig. 5). Additionally, most respondents believed that student
instructors provided quality answers, with 88.6\% indicating that
their questions were usually answered well (Fig. 6).

Impact of student teachers on CSE retention was mostly positive. Students
were asked whether student teachers had a positive impact, negative
impact, or no impact on their intentions to continue in, or switch
to, a CSE major. In total, 70.7\% of responses indicated that they
were swayed towards a CSE major by student teachers, with 28.6\% professing
no impact on their educational goals, and only 0.7\% claiming a negative
impact.

Former and graduating student teachers were asked to evaluate the
ways in which the SPL program affected their technical skills, and
100\% of respondents indicated that their coding and programming concept
skills were improved by participation in the program. Respondents
also overwhelmingly indicated that their interview skills were improved
significantly by serving as a student teacher, with 93.1\% of respondents
asserting that the experience increased their confidence in answering
technical questions. 

Free responses from the former and graduating student teachers were
mostly positive. A total of 62.1\% of respondents included praise
for the fact that as student teachers they were able to learn many
important computer science related social skills while they taught
others. Additionally, the vast majority of respondents expressed satisfaction
with the program, with 96.6\% of responses indicating an overall positive
experience.
\begin{center}
\begin{figure}
\begin{centering}
\includegraphics[scale=0.6]{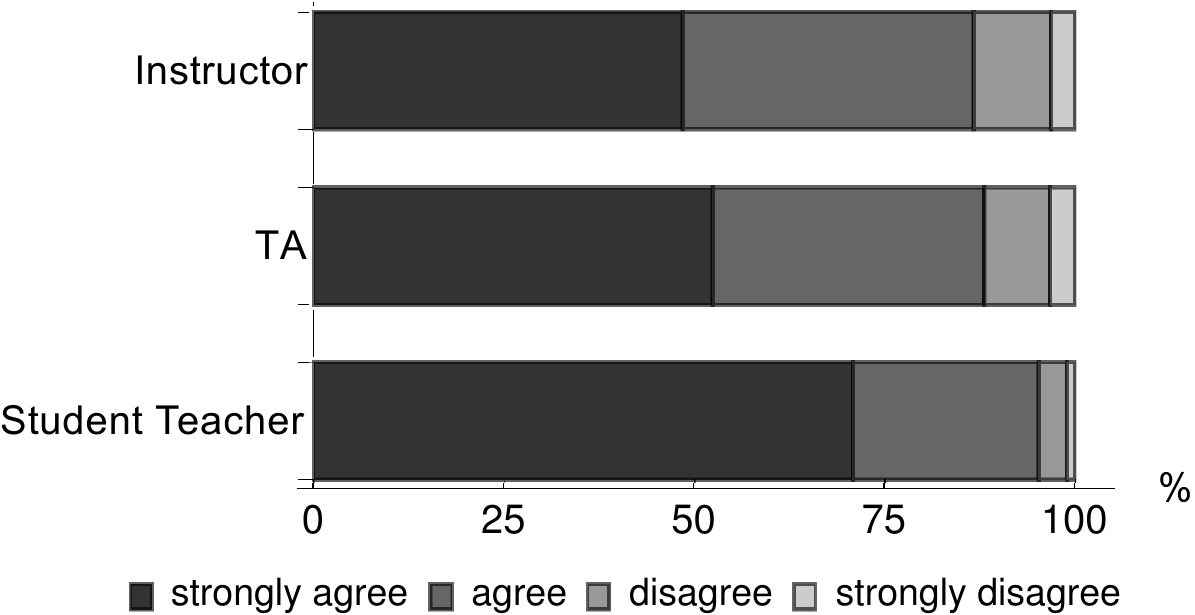}
\par\end{centering}
\centering{}\caption{Level of Comfort}
\end{figure}
\par\end{center}

\begin{center}
\begin{figure}
\begin{centering}
\includegraphics[scale=0.55]{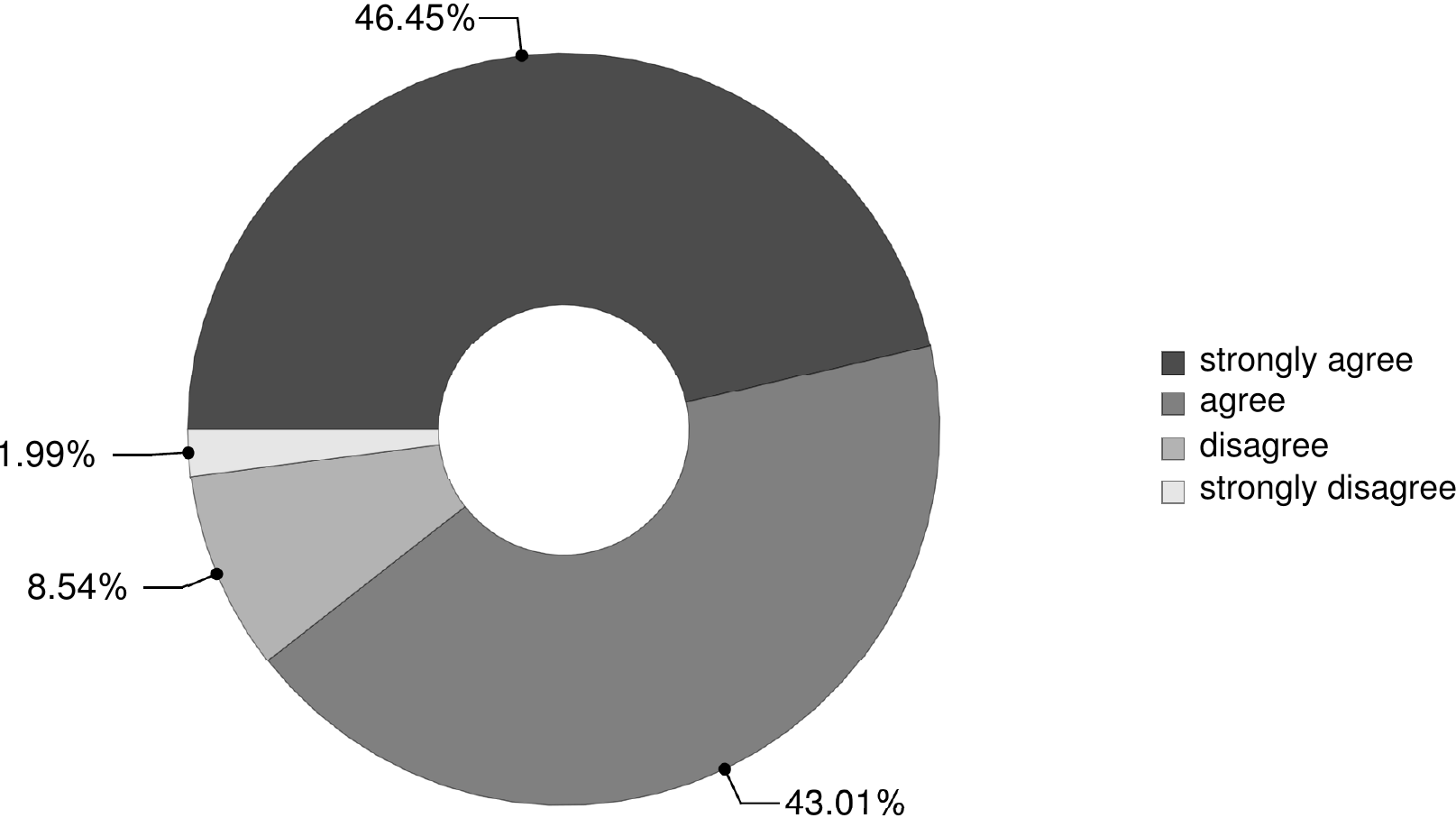}
\par\end{centering}
\centering{}\caption{Did the program help you to get a better grade?}
\end{figure}
\par\end{center}

\begin{center}
\begin{figure}
\begin{centering}
\includegraphics[scale=0.55]{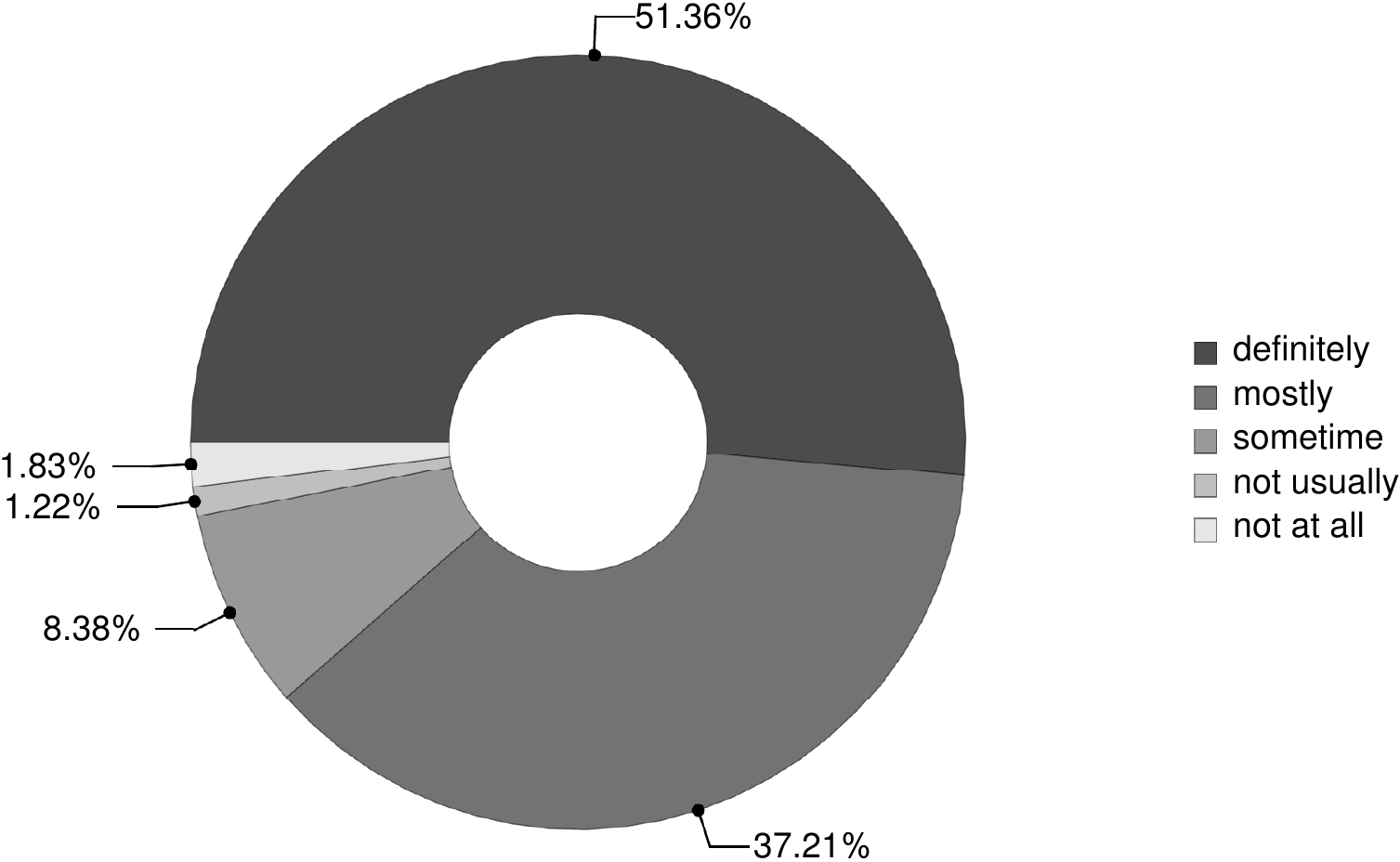}
\par\end{centering}
\centering{}\caption{Were your questions answered well?}
\end{figure}
\par\end{center}

\section{Discussion}

Computer Science education presents many unique challenges, and creating
an environment in which students are willing, capable, and interested
in learning can be a very difficult task, see \cite{MAl13}. The SPL
program has shown that it is successful in creating an environment
in which students are comfortable seeking help, and that the assistance
provided by the student teachers has, in the students\textquoteright{}
estimation, greatly increased students ability to succeed in Computer
Science courses. Additionally, results show that, in the majority
of cases, the program inclines students towards staying in, or switching
to, a CSE major.

In addition to the improvements experienced by students who received
help from the SPL program, benefits appear to have been reaped by
the high-performing students serving as student teachers. Improved
interview and communication skills were indicated by an overwhelming
majority of former student teachers, and an improvement in programming
and abstract thinking skills were enjoyed ubiquitously. This data
supports the hypothesis that SPL is a program that emphasizes bi-directional
learning, in which both the student and student teacher are able to
benefit from the process of teaching. Student teachers also expressed
general satisfaction with the program. Many respondents indicated
that they felt they had succeeded in providing effective mentorship
to students, in areas directly related to programming as well as in
other related skills.

The historical data from 2007\textendash 2008 regarding student comfort
in asking for help from student teachers agrees with the more recent
data. In addition, the data concerning GPA provides concrete evidence
for our contention that student grades are impacted positively by
the SPL program. This data establishes a clear increase in GPA for
students in lower level courses who took advantage of the program,
and shows that the SPL program is capable of meeting one of its primary
goals which is increasing student success in CSE courses.

\section{Conclusion}

The SPL program in the CSE department at Texas A\&M University has
been successfully providing deep learning benefits to all participants
for nearly a decade, and we believe that it not only provides an additional
source of help to students who have needed it, but also motivates
and inspires students to learn, creates an effective learning culture,
and builds stronger social interaction and cognitive development.
The program continues to develop based on the feedback we receive
from both students and student teachers. It has increased in size
since 2009, and continues to grow as it facilitates success in the
undergraduate CSE bodies. The program is not perfect, and cannot ensure
the success of every student, because those who do not wish to put
forth the effort and seek help see no benefits from the program. However,
we are confident that the methods used are effective and that the
implementation of Structured Peer Learning at the department is on
the right path.\enlargethispage{\baselineskip}

\section*{Acknowledgments}

The authors would like to thank the Texas A\&M University Computer
Science and Computer Engineering department for its continued support,
as well as the TAMU CSCE Student Teachers for their untiring and selfless
dedication to furthering the education of their peers. 

\bibliographystyle{acm}
\bibliography{tleyk_2017}

\begin{thebibliography}{10}

\bibitem{Bon96}
{\scshape Bonwell, C.~C.}
\newblock Enhancing the lecture: Revitalizing the traditional format.
\newblock {\em New Directions for Teaching and Learning 67\/} (1996), 31--44.

\bibitem{CLi11}
{\scshape Chen, C., and Liu, C.~C.}
\newblock A case study of peer tutoring program in higher education.
\newblock {\em Research in Higher Education 11\/} (2011), 1--10.

\bibitem{Den00}
{\scshape Denning, P.~J.}
\newblock {C}omputer {S}cience: The discipline.
\newblock Encyclopedia of Computer Science, 2000.
\newblock Archived from the original (PDF) on 2006-05-25.

\bibitem{Fal01}
{\scshape Falchikov, N.}
\newblock {\em Learning Together: Peer Tutoring in Higher Education}.
\newblock RoutledgeFalmer, New York, NY, 2001.

\bibitem{Gol76}
{\scshape Goldschmid, B.}
\newblock Peer teaching in higher education: A review.
\newblock {\em Higher Education 5\/} (1976), 9--33.

\bibitem{LRK14}
{\scshape Lazowska, E., Roberts, E., and Kurose, J.}
\newblock Tsunami or sea change? {R}esponding to the explosion of student
  interest in {C}omputer {S}cience.
\newblock Slides for CRA Conference at Snowbird, July 2014.
\newblock Also Slides for NCWIT 10th Anniversary Summit in May 2014.

\bibitem{MAl13}
{\scshape Mhashi, M.~M., and Alakeel, A.}
\newblock Difficulties facing students in learning computer programming skills
  at {T}abuk {U}niversity.
\newblock {\em Recent Advances in Modern Educational Technologies\/} (2013),
  15--24.

\bibitem{NWi66}
{\scshape Newcomb, T.~M., and Wilson, E.~K.}, Eds.
\newblock {\em College peer groups: problems and prospects for research\/}
  (1966), Aldine Publishing Company.

\bibitem{Pri04}
{\scshape Prince, M.}
\newblock Does active learning work? {A} review of the research.
\newblock {\em Journal of Engineering Education 93}, 3 (2004), 223--231.

\bibitem{RHT01}
{\scshape Ramaswamy, S., Harris, I., and Tschirner, U.}
\newblock Student peer teaching: An innovative approach to instruction in
  science and engineering education.
\newblock {\em Journal of Science Education and Technology 10}, 2 (2001),
  165--171.

\bibitem{Str09}
{\scshape Stroustrup, B.}
\newblock Programming in an undergraduate {CS} curriculum.
\newblock In {\em Proceedings of the 14th {W}estern {C}anadian {C}onference on
  {C}omputing {E}ducation\/} (New York, NY, 2009), R.~Brouwer, D.~Cukierman,
  and G.~Tsiknis, Eds., ACM, pp.~82--89.

\bibitem{Top96}
{\scshape Topping, K.~J.}
\newblock The effectiveness of peer tutoring in further and higher education: A
  typology and review of the literature.
\newblock {\em Higher Education 32\/} (1996), 321--345.

\end{thebibliography}

\end{document}